\documentstyle[aps,prl,preprint,epsf,floats]{revtex}

\newcommand{\beq}{\begin{equation}}
\newcommand{\eeq}{\end{equation}}
\newcommand{\bqa}{\begin{eqnarray}}
\newcommand{\eqa}{\end{eqnarray}}

\def\square{\vcenter{\vbox{\hrule height.4pt
          \hbox{\vrule width.4pt height8pt
          \kern8pt\vrule width.4pt}\hrule height.4pt}}}

\begin{document}

\preprint{
\vbox{\halign{&##\hfil\cr
&ITF-uu-00/40\cr
        & hep-ph/yymmnn \cr
&\today\cr }}}

\title{Ground State of a Trapped Bose-Einstein Condensate in Two Dimensions;
Beyond the Mean-field Approximation }
\author{Jens O. Andersen
}
\address{Institute for Theoretical Physics, University of Utrecht,
Princetonplein 5, 3584 CC Utrecht, The Netherlands}

\author{H{\aa}rek Haugerud}
\address{Oslo University College, Faculty of Engineering, Cort Adelers gate 30,
0254 Oslo, Norway}
\maketitle

\begin{abstract}
{\footnotesize 
We consider the ground state of a trapped Bose-Einstein condensate
in two dimensions. In the mean-field approximation, the ground state 
density profile
satisfies the Gross-Pitaevskii equation.
We compute the leading quantum corrections to the density profile
to second order in an expansion around the Thomas-Fermi limit.
By summing the ladder diagrams,
we are generalizing Schick's result for the ground state energy
of a homogeneouns Bose gas to the case of a trapped Bose gas.
}
\end{abstract}

\newpage
\section{Introduction}
The remarkable achievement of Bose-Einstein condensation (BEC) of alkali
atoms in harmonic traps~\cite{bec1}--$\!\!\!$\cite{bec3} 
has created an enormous interest in the properties of dilute Bose gases
(see e.g. Ref.~\cite{string} and references therein).
One of the basic questions is the condensate density profile as a function of
temperature. At zero temperature and in the mean-field approximation, the
condensate density satisfies the Gross-Pitaevskii equation.
Until recently, the condensates were so dilute that mean-field theory
gives a satisfactory description of the experiments.
However, there are corrections from 
quantum fluctuations around the mean field and their relative importance
grow like the gas parameter
$\sqrt{\rho a^3}$, where $a$ is the S-wave scattering length and $\rho$
is the density. 
The sign and value 
of the scattering length are determined by the position of the least bound
molecular bound states. 
The position of the bound states can be manipulated by applying external
fields.
In recent experiments, Cornish {\it et al}. were able to vary
the scattering length
$a$ over a large range by applying a strong external magnetic field
and exploiting the existence of a Feshbach resonance at $B\sim 155\;G$.
Values for $\sqrt{\rho a^3}$ up to approximately
0.1 were obtained and should be sufficiently large to see deviations
from the mean field in experiments.
Hence it becomes 
important to be able to calculate the effects of quantum fluctuations
around the mean field in a systematic way.
Such an approach was developed in Ref.~\cite{eric}, where the effects of
quantum fluctuations on the ground state of a Bose-Einstein
condensate in three dimensions were calculated. In 
Ref.~\cite{jenseric}, the result was extended to arbitrary 
time-independent states, including vortices.
The method is a combination of the Hartree-Fock approach and the Thomas-Fermi
approach. The Hartree-Fock method includes all the leading quantum corrections
to the mean-field
equation for the density. The resulting equation
is an integral equation, which is reduced
to a local differential equation by applying a gradient expansion around the
Thomas-Fermi limit.

In a new paper, Blume and and Greene~\cite{qc}, use a diffusion Monte Carlo
method to calculate the ground state energy for different two-body
potentials that generate the same value for the scattering length $a$.
It turns out that the ground state energy
is independent of the actual form of the potential; it only depends on $a$.
Moreover, it differs significantly from the ground state energy obtained
from solving the Gross-Pitaevskii equation. Including the 
leading quantum corrections to zeroth order in a gradient expansion,
the agreement with the result obtained by Blume and 
Greene~\cite{qc} improves significantly.

It is well known that a Bose-Einstein condensate
in a two-dimensional homogeneous Bose gas 
only exists at $T=0$~\cite{hohen}. 
At any finite temperature, the phase fluctuations 
destroy the condensate. This reflects the Mermin-Wagner theorem 
stating that there is no spontaneous breakdown of a continuous symmetry
in a homogeneous system in two dimensions at finite temperature.

The ground state energy density of a two-dimensional homogeneous Bose gas
was first calculated by Schick~\cite{schiff}.
By summing up the ladder diagrams contributing to the
chemical potential, he was able to obtain the leading correction to the
mean field result.
The result has later been derived by several authors~\cite{fis,rg2}. 
Very recently, a formal proof of the result by Schick was given by
Lieb and Yngvason~\cite{lieb1,lieb2}. 

The Mermin-Wagner theorem does not apply to an inhomogeneous system
and a condensate may exist.
Bagnato and Kleppner~\cite{klep} showed 
the possibility of BEC for $T<T_c\approx\sqrt{N}\hbar\omega$ 
(where $N$ is the number of particles and $\omega$ is the trap frequency)
in a two-dimensional ideal gas confined by a harmonic potential.
In this temperature
range, the phase fluctuations on the scale of the size of
a trapped gas is
negligible, and there is a true condensate.
In Refs.~\cite{jason,petrov}, it was shown that the possibility of BEC
exists also when local two-body interactions are included.

The ground state properties of a trapped Bose gas in two dimensions have been 
studied recently by a number of authors. In Refs.~\cite{torhaarek,adhi}, the
Gross-Pitaevskii equation was solved numerically. 
However, the Gross-Pitaevskii equation is a mean-field
equation and receives quantum corrections. 
In Refs.~\cite{lieb1,lieb2,kolo}, it is
argued that the correct equation for the density
can be derived from an energy functional that includes
the corrections obtained by Schick~\cite{schiff}.

However, since a trapped Bose gas is an inhomogeneous system, the 
Gross-Pitaevskii equation also receives corrections that involve gradients
of the density. In the present paper, we calculate the leading
quantum corrections to the
Gross-Pitaevskii 
equation to second order in the gradient expansion around the Thomas-Fermi
limit. 
We also apply the $T$-matrix approximation to sum up the ladder diagrams.
The resulting equation 
extends the result in Refs.~\cite{lieb1,lieb2,kolo}.

The plan of the paper is as follows.
In section II, we review the perturbative framework developed in 
Refs.\cite{eric} to calculate the leading quantum corrections to 
the the Gross-Pitaevskii equation.
In section III, we calculate the self-consistent one-loop
corrections to second order in the gradient expansion.	
In section IV, we briefly discuss the summation of the ladder diagrams
and derive an equation that takes this summation into account.
Finally, we summarize our results in section V.
Calculational details are included in
two appendices.
\section{Perturbative Framework}
In this section, we discuss the perturbative framework developed in 
Ref.~\cite{eric} to calculate the leading 
effects on the ground state
from quantum fluctuations around the mean field.

The action is
\bqa\nonumber
S[\psi]&=&\int dt\;
\Bigg(\int d^2x\;
\psi^*({\bf x},t)
\left[
i\hbar{\partial\over\partial t}
+{\hbar^2\over2m}\nabla^2+\mu-V({\bf x})
\right]\psi({\bf x},t)
\\ 
&&
\label{ac}
-{1\over2}\int d^2x\int d^2x^{\prime}\;
\psi^*({\bf x},t)\psi^*({\bf x}^{\prime},t) V_0({\bf x}-{\bf x}^{\prime})
\psi({\bf x},t)\psi({\bf x}^{\prime},t)
\Bigg)\;.
\eqa

$\psi^*({\bf x},t)$ is complex field operator that creates a boson at the
position ${\bf x}$,
$V({\bf x})$ is the trapping potential, $V_0({\bf x})$ 
is the two-body potential. In the following, we set $\hbar=2m=1$.
Factors of $\hbar$ and $2m$ can be reinserted using dimensional analysis.

The action~(\ref{ac}) is symmetric under a phase transformation
\bqa
\psi({\bf x},t)\rightarrow e^{i\alpha}\psi({\bf x},t)\;.
\eqa

This $U(1)$-symmetry ensures that the density $\rho$ and current 
density ${\bf j}$
satisfy the continuity equation
\bqa
\dot{\rho}+\nabla\cdot{\bf j}=0\;.
\eqa

In the ground state, the current density ${\bf j}$ vanishes identically and
the condensate has a constant phase. The $U(1)$-symmetry can then be used to
make the condensate real everywhere.

In the following, we approximate the two-body interaction 
$V_0({\bf x}-{\bf x}^{\prime})$ by a 
delta function with strength $g$:
\bqa
\label{ac2}
V_0({\bf x}-{\bf x}^{\prime})=g\delta({\bf x}-{\bf x}^{\prime})\;.
\eqa

We can then trivially integrate over ${\bf x}^{\prime}$, and the
action~(\ref{ac}) simplifies to
\bqa
\label{act}
S[\psi]&=&\int dt\;
\int d^2x\;
\psi^*
\left[
i
{\partial\over\partial t}
+
\nabla^2+\mu-V
\right]\psi
-{1\over2}
g\Big(\psi^*\psi\Big)^2
\;.
\eqa

The quantum field theory defined by the action~(\ref{act})
has ultraviolet divergences that must be removed by renormalization of
$\mu$ and $g$. 
There is also an ultraviolet divergence in the expression for the density
$\rho$. This divergence can be removed by adding a counterterm $\delta\rho$.
Alternatively, one can eliminate the divergences associated with
$\mu$ and $\rho$ by a normal-ordering prescription of the fields 
in Eq.~(\ref{act}). The coupling constant is renormalized in the usual way
by replacing the bare coupling with the physical one.

If we use a simple momentum cutoff $M$ to cut off the
ultraviolet 
divergences in the loop integrals, there will be terms proportional to
$M^p$, where $p$ is a positive integer.
There are also terms that are proportional to $\log M$.
The coefficients of the power divergences depend on the regularization
method and are therefore artifacts of the regulator.
On the other hand, the coefficients of $\log(M)$ are independent of the
regulator and they therefore represent real physics.
In this paper, we use dimensional regularization 
to regulate both infrared and ultraviolet divergences.
In dimensional regularization, one calculates the loop
integrals in $d=2-2\epsilon$ dimensions for values of $\epsilon$ where
the integrals converge. One then analytically continues back to $d=2$
dimensions.
With dimensional regularization, an arbitrary renormalization scale $M$
is introduced. This scale can be identified with the simple momentum
cutoff mentioned above.
An advantage of dimensional regularization is that it 
automatically sets power 
divergences to zero, while logarithmic divergences
show up as poles in $\epsilon$.
In two dimensions, the one-loop 
counterterms for the chemical potential $\mu$ and the density $\rho$
are quadratic ultraviolet
divergences, while the one-loop counterterm for the coupling constant
$g$ is a logarithmic ultraviolet divergence.

We next parameterize the quantum field $\psi$
in terms of a time-independent
condensate $v$ and a quantum fluctuating field $\tilde{\psi}$:
\bqa
\psi=v+\tilde{\psi}\;,
\eqa
where the condensate $v$ satisfies
\bqa
v=\langle\psi\rangle\;.
\eqa
Here and in the following, $\langle A\rangle$ denotes expectation value
of the operator $A$ in the ground state.
Thus the expectation value of $\tilde{\psi}$ vanishes.
The fluctuating field can be written in terms of two real fields:
\bqa
\label{split}
\tilde{\psi}={1\over\sqrt{2}}\left(\psi_1+i\psi_2\right)\;.
\eqa
Substituting Eq.~(\ref{split}) into Eq.~(\ref{ac2}), the action can 
be decomposed into three terms
\bqa
\label{terms}
S[\psi]=S[v]+S_{\rm free}[\psi_1,\psi_2]
+S_{\rm int}[v,\psi_1,\psi_2]\;.
\eqa
$S[v]$ is the classical action
\bqa
S[v]=\int dt\int d^2x
\left[\left(\mu -V\right)v^2-{1\over2}gv^4+v\nabla^2v\right]\;,
\eqa
while the free part of the action is
\bqa
\label{free}
S_{\rm free}[\psi_1,\psi_2]=\int dt\int d^2x
\left[
{1\over2}\left(\dot{\psi}_1\psi_2-\psi_1\dot{\psi_2}\right)
+{1\over2}
\psi_1\left(\nabla^2-\Lambda^2\right)\psi_1+{1\over2}
\psi_2
\nabla^2\psi_2
\right]\;.
\eqa
The interaction part of the action is
\bqa\nonumber
S_{\rm int}[v,\psi_1,\psi_2]&=&
\int dt\int d^2x\left[
\sqrt{2}T\psi_1+{1\over2}
X\psi_1^2+{1\over2}Y
\psi_2^2
+{1\over\sqrt{2}}Z\psi_1\left(\psi_1^2+\psi_2^2\right)
\right.\\ 
\label{inter}
&&\left.-{1\over8}g\left(\psi_1^2+\psi_2^2\right)^2
\right]\;.
\eqa
 The sources in Eq.~(\ref{inter}) are
\bqa
T&=&\left[\mu-V-gv^2\right]v+
\nabla^2v\;,\\
X&=&\Lambda^2+
\left[\mu-V-3gv^2\right]\;,\\
Y&=&
\left[\mu-V-gv^2\right]\;,\\
Z&=&-gv\;.
\eqa
 Note that we have added and subtracted an arbitrary 
term ${1\over2}\Lambda^2\psi_1^2$
in the action. 
effects are subtracted out at higher orders through the source $X$.
By a judicious choice of $\Lambda$, one
can ensure that $X$ can be treated as a perturbation in the same way as
the other sources and thus simplify calculations. 
We return to the choice of $\Lambda$ in the next section.

The propagator that corresponds to the free action $S_{\rm free}
[\psi_1,\psi_2]$
in Eq.~(\ref{free}) is
\bqa
\label{prop}
D(\omega,k)&=&\frac{i}{\omega^2
-\epsilon^2(k)+i\epsilon}\left(\begin{array}{cc}
k^2&-i\omega \\
i\omega&\epsilon^2(k)/k^2
\end{array}\right)\;.
\eqa
Here ${\bf k}$ is the wavevector, $\omega$ is the frequency, and
$\epsilon(k)$ is the dispersion relation for the Bogoliubov modes:
\bqa
\epsilon(k)=k\sqrt{k^2+\Lambda^2}\;.
\eqa
The dispersion relation is gapless, which 
reflects the spontaneous breakdown of the $U(1)$-symmetry
(Goldstone's theorem). The dispersion
is quadratic for large wavevectors and is that of a free nonrelativistic 
particle. The propagator is defined with an $i\epsilon$ prescription
in the usual way. The diagonal parts of the propagator are denoted
by a solid and a dashed line, respectively. The off-diagonal parts
are denoted by a solid-dashed and dashed-solid line, respectively.

The quantum field theory defined by Eq.~(\ref{act}) has been decomposed
into  free and interacting parts. The quantum
loop expansion is then an expansion
in the dimensionless coupling constant $g$. The Hartree-Fock approximation
includes all effects to leading order in $g$.

The field equation for $\psi_1$
(and $\psi_2$)
is obtained by varying the action~(\ref{terms}). 
Taking the expectation value of the equation for $\psi_1$, we obtain the
tadpole equation
\bqa
\label{t1}
0=T+{3\over2}Z\langle\psi_1^2\rangle+{1\over2}Z\langle\psi_2^2\rangle
-{1\over2\sqrt{2}}g\langle\psi_1(\psi_1^2+\psi_2^2)\rangle\;.
\eqa
The density can be written as
\bqa\nonumber
\rho&=&\langle\psi^*\psi\rangle \\
\label{rho}
&=&v^2+{1\over2}\langle\psi_1^2\rangle+{1\over2}\langle\psi_2^2\rangle\;,
\eqa
To zeroth order in the loop expansion
one neglects the expectation values in Eq.~(\ref{t1}) and the tadpole equation
reduces to $T=0$. 
We also neglect the expectation values in Eq.~(\ref{rho}) and we
have $\rho=v^2$. The tadpole equation then reduces to the 
Gross-Pitaevskii equation for the density profile:
\bqa
\label{gross}
\left[
\mu-V
-
g\rho\sqrt{\rho}\right]
+\nabla^2\sqrt{\rho}
=0\;.
\eqa
The last term in Eq.~(\ref{t1}) only contributes at second order 
and beyond in the loop
expansion. Thus to leading order in the quantum fluctuations,
Eq.~(\ref{t1}) reduces to
\bqa
\label{t2}
0=T+{3\over2}Z\langle\psi_1^2\rangle+{1\over2}Z\langle\psi_2^2\rangle\;.
\eqa
Eq.~(\ref{t2}) is referred to as the semiclassical tadpole 
equation~\cite{eric}.

The sources $X$ and $Y$ depend on the condensate $v$. In order to obtain
an equation for the density $\rho$, we invert Eq.~(\ref{rho}) so that we can 
eliminate $v$ in favour of $\rho$ in Eq.~(\ref{t2}).
The expectation values in Eq.~(\ref{rho}) are 
correction terms arising from quantum fluctuations around the mean field.
These terms are suppressed by powers
of $g$ compared to the terms in the classical equations.
Since we are only interested in the leading quantum effects, we can
invert Eq.~(\ref{rho})  
and expand it to first order in the expectation values:
\bqa
\label{v}
v=\sqrt{\rho}-{1\over4\sqrt{\rho}}\langle\psi_1^2+\psi_2^2\rangle\;.
\eqa

We can also derive expressions for the gradients of $v$ by 
differentiating Eq.~(\ref{v}):
\bqa
\nabla v&=&
{1\over4\rho}\left[
4\rho\nabla\sqrt{\rho}+\nabla\sqrt{\rho}\langle\psi_1^2+\psi_2^2\rangle
-\sqrt{\rho}\;\nabla\langle\psi_1^2+\psi_2^2\rangle
\right]\;,\\ \nonumber
\nabla^2v&=&{1\over4\rho^{3/2}}
\left[
4\rho^{3/2}\nabla^2\sqrt{\rho}
+\left[\sqrt{\rho}\;\nabla^2\sqrt{\rho}-2\left(\nabla\sqrt{\rho}\right)^2\right]\langle
\psi_1^2+\psi_2^2\rangle\right. \\
\label{ddv}
&&\left.+2\sqrt{\rho}\;\nabla\sqrt{\rho}\cdot\nabla\langle\psi_1^2+\psi_2^2\rangle
-\rho\nabla^2\langle\psi_1^2+\psi_2^2\rangle
\right]\;.
\eqa

Substituting
the expressions (\ref{v}) for $v$ and~(\ref{ddv}) for $\nabla^2v$, the
semiclassical tadpole equation~(\ref{t2}) becomes
\bqa\nonumber
0&=&\left[\mu-V-g\rho\right]\sqrt{\rho}
+\nabla^2\sqrt{\rho}
-g\sqrt{\rho}
\langle\psi_1^2\rangle
+{1\over2\rho^{3/2}}
\left[\sqrt{\rho}\nabla^2\sqrt{\rho}-\left(\nabla\sqrt{\rho}\right)^2\right]
\langle\psi_1^2+\psi_2^2\rangle
\\ \nonumber
&&
+{1\over2\rho}\nabla\sqrt{\rho}\cdot\nabla\langle\psi_1^2+\psi_2^2\rangle
-{1\over4\sqrt{\rho}}\nabla^2
\langle\psi_1^2+\psi_2^2\rangle \\
&&
\label{last}
-{1\over4\sqrt{\rho}}
\left(\mu-V-g\rho
+{1\over\sqrt{\rho}}\nabla^2\sqrt{\rho}\right)
\langle\psi_1^2+\psi_2^
2\rangle\;.
\eqa
The last line in Eq.~(\ref{last}) is proportional to the classical equation of
motion. The corrections to the classical equation of motion are of order $g$
and so are the expectation values of the quantum fields.
Thus the last line is 
second order in quantum corrections and will therefore be dropped
in the following.

\section{Gradient Expansion}
The expectation values in Eq.~(\ref{t2}) are functionals of the sources $X$ 
and $Y$. These functionals involve an arbitrary number of insertions of 
the sources in the loop diagrams for $\langle\psi_1^2\rangle$
and $\langle\psi_2^2\rangle$.
They are non-local since the positions of the sources are 
integrated over. Thus the tadpole is an integral equation.

Instead of solving the non-local integral equation~(\ref{last}), 
we would like
to derive a local differential equation by expanding the expectation
values $\langle\psi_1^2\rangle$ and $\langle\psi_2^2\rangle$
in powers of the sources and their
gradients. The Thomas-Fermi expansion is an expansion in powers of 
$\xi/R$, where $\xi=1/\sqrt{g\rho}$ 
is the local coherence length and $R$ is the
length scale for significant changes in the density.
It is possible to expand the expectation values in Eq.~(\ref{t2}) 
in powers of gradients
of the sources, if the expectation values receive significant 
contributions only
from modes with wavelengths that are of the order of the
coherence length or less. This can be ensured by introducing an infrared
cutoff that eliminates the contribution from wavelengths larger than the
coherence length. We use dimensional regularization to 
guarantee that the effects of these modes are eliminated.
If the results depend on the infrared cutoff, it
indicates a sensitivity to length scales much larger than the coherence
length~\cite{eric}.

We can also expand the expectation values in powers of $X$ and $Y$ if
the sources are at least either first order in the gradient expansion 
or in the coupling constant $g$. 
This is generally not correct, but it 
can be made true at a specific point
${\bf x}_0$ by a clever choice of the parameter $\Lambda$~\cite{eric}.
Since the expectation values of the quantum fields are of order $g$
and we are interested in the leading quantum effects, 
we can use the classical equation of motion, $T=0$, to simplify the sources:
\bqa
X({\bf x}_0)&=&\Lambda^2-2gv^2-{1\over v}\nabla^2v\;,\\
Y({\bf x}_0)&=&-{1\over v}\nabla^2v\;.
\eqa 
The source $Y({\bf x}_0)$ is already second order in the 
gradient expansion. We can also
make $X({\bf x}_0)$ 
second order in the gradient expansion by the following choice~\cite{eric} 
for the arbitrary parameter $\Lambda$
\bqa
\label{choice}
\Lambda^2=2g\rho({\bf x}_0)\;.
\eqa
The difference between $\rho$ and $v^2$ is higher order in $g$
and can be neglected. 
Any other choice of $\Lambda$ which is equal to 
Eq.~(\ref{choice}) to leading order
in $g$ and in the gradient expansion, results in the same final equation
for the density.

\begin{figure}[htb]
\begin{center}

\begin{picture}(0,0)%
\epsffile{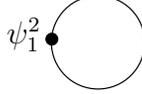}%
\end{picture}%
\setlength{\unitlength}{1973sp}%
\begingroup\makeatletter\ifx\SetFigFont\undefined%
\gdef\SetFigFont#1#2#3#4#5{%
  \reset@font\fontsize{#1}{#2pt}%
  \fontfamily{#3}\fontseries{#4}\fontshape{#5}%
  \selectfont}%
\fi\endgroup%
\begin{picture}(1728,1200)(1051,-1730)
\put(1051,-1111){\makebox(0,0)[lb]{\smash{\SetFigFont{12}{14.4}{\familydefault}{\mddefault}{\updefault}$\psi_{1}^{2}$}}}
\end{picture}

\vspace{2mm}
\caption[a]{Feynman diagram needed for the calculation of the 
coefficient in $a_0$ Eq.~(\ref{p1})}
\label{ex}
\end{center}
\end{figure}

The expectation values in Eq.~(\ref{last})
are now expanded in powers of $X$ and $Y$ and their
gradients around the point ${\bf x}_0$
\bqa
\label{p1}
\langle\psi_1^2\rangle&=&a_0+a_1X+a_2Y+a_3\nabla^2X+a_4X^2
+a_5\left(\nabla X\right)^2+...\;,\\
\langle\psi_2^2\rangle&=&b_0+b_1X+b_2Y+b_3\nabla^2X+b_4X^2
+b_5\left(\nabla X\right)^2+...\;.
\label{p2}
\eqa
The expansions of $\langle\psi_1\rangle$ and $\langle\psi_2\rangle$
include all rotationally invariant terms built from the sources and 
their derivatives. In Fig.~\ref{ex}, the diagram that we need to calculate
in order to obtain the coefficient $a_0$ appearing in
Eq.~(\ref{p1}). This calculation is sketched
in Appendix B.
\vspace{2mm}

We will also need to calculate the gradients of $\langle\psi_1^2\rangle$
and $\langle\psi_2^2\rangle$ in powers of the sources and their derivatives.
Adding Eqs.~(\ref{p1}) and~(\ref{p2}) and differentiating, we obtain
\bqa
\label{p3}
\nabla\langle\psi_1^2+\psi_2^2\rangle&=&
(a_1+b_1)\nabla X+...
\;,\\
\label{p4}
\nabla^2\langle\psi_1^2+\psi_2^2\rangle&=&
(a_1+b_1)\nabla^2X+
2(a_4+b_4)\left(\nabla X\right)^2+...
\;.
\eqa

We next eliminate the dependence on $v$ of the sources in favour of the
density $\rho$. Since the sources only appear in the expectation values
that are of order $g$, we can neglect the expectation values in 
Eqs.~(\ref{v})--(\ref{ddv}):
\bqa
v({\bf x}_0)&=&\sqrt{\rho}\;,\\
\nabla v({\bf x}_0)&=&
\nabla\sqrt{\rho}
\;,\\
\nabla^2v({\bf x}_0)&=&
\nabla^2\sqrt{\rho}
\;.
\eqa
The resulting expressions for the sources and their derivatives
then become
\bqa
\label{s1}
X({\bf x}_0)&=&-{\nabla^2\sqrt{\rho}\over\sqrt{\rho}}\;,\\
\nabla X({\bf x}_0)&=&-4g\sqrt{\rho}\;\nabla\sqrt{\rho}\;,\\
\nabla^2X({\bf x}_0)&=&-4g\left[\sqrt{\rho}\;
\nabla^2\sqrt{\rho}+\left(\nabla\sqrt{\rho}\right)^2
\right]\;, \\
\label{s3}Y({\bf x}_0)&=&
-{\nabla^2\sqrt{\rho}\over\sqrt{\rho}}\;.
\eqa
Using Eqs.~(\ref{p1})--(\ref{p4}) for the expectation values
and the expressions~(\ref{s1})--(\ref{s3}) for the sources, 
the semiclassical tadpole equation reduces to
\bqa\nonumber
0&=&\left(\mu-V-g\rho\right)\sqrt{\rho}+
\nabla^2\sqrt{\rho}
-a_0g\sqrt{\rho} \\ \nonumber
&&+{1\over2}\left(a_0+b_0+(2a_1+b_1+a_2)\Lambda^2
+2a_3\Lambda^4
\right){\nabla^2\sqrt{\rho}\over \rho}\\ 
\label{eqcoef}
&&-{1\over2}\left(a_0+b_0+(a_1+b_1)\Lambda^2
-2(a_3-2a_4-2b_4)\Lambda^4+4a_5\Lambda^6
\right){\left(\nabla\sqrt{\rho}\right)^2\over 
\rho\sqrt{\rho}}\;.
\eqa
The coefficients $a_i$ and $b_i$ were calculated in Ref.\cite{eric}.
They can be expressed in terms of the integral $I_{m,n}$, which is defined
in Appendix A. 
We list them in Appendix B for convenience.
The coefficients $a_0$ and $b_0$ are quadratically ultraviolet divergent,
while $a_1$, $a_2$, $b_1$, and $b_2$ are logarithmically ultraviolet
divergent. The remaining coefficients are ultraviolet finite.
Note also that the infrared divergences in the individual coefficients cancel
in the sum. The fact that the dependence on the infrared cutoff cancels, 
ensures that there is no sensitivity to length scales much larger than the
coherence length to leading order in the gradient expansion.

The counterterm needed to cancel the ultraviolet
divergences in Eq.~(\ref{eqcoef}) is~\cite{berg}
\bqa
\delta g&=&{g^2\over8\pi\epsilon}\;.
\eqa
Using the values for the coefficients listed in Appendix B, we obtain
\bqa
\label{final}
\left[
\nabla^2+\mu-V\right]\sqrt{\rho}
-g\rho\sqrt{\rho}-{g^2\over8\pi}\Bigg\{1+\log
\left[{g\rho\over2M^2}\right]\Bigg\}\rho\sqrt{\rho}
-{g\over12\pi}
\nabla^2\sqrt{\rho}
=0\;.
\eqa
Note in particular that the coefficient 
of $\left(\nabla\sqrt{\rho}\right)^2$ in Eq.~(\ref{eqcoef})
vanishes. This is not the case in three dimensions~\cite{eric} and we
have no explanation for this cancellation.
Eq.~(\ref{final}) has been derived at a specific point ${\bf x}_0$.
Since the point ${\bf x}_0$ is arbitrary, Eq.~(\ref{final})
must hold everywhere.

The logarithmic term in Eq.~(\ref{final}) can be obtained from the
one-loop result for the ground state energy density ${\cal E}$
for a homogeneous Bose gas, 
obtained in Refs.~\cite{loz,finn},
by differentiating ${\cal E}$ with respect to $\sqrt{\rho}$. 
The last term in Eq.~(\ref{final}) is a new result.

This renormalization scale is completely arbitrary and physical quantities
such as the density profile of the ground state must be independent of it.
The requirement that physical quantities be independent of $M$ can 
be expressed in terms of renormalization group equations for the coupling
constants in the Lagrangian. 
The coupling constant $g$ in Eq.~(\ref{act}) 
satisfies
\bqa
\label{rg2}
M{\!\!d\over dM}\;g&=&\beta(g)\;,
\eqa
where the $\beta$-function is a polynomial in $g$.
Normally, these functions are
known only up to a certain order in the quantum loop
expansion. At the one-loop level 
$\beta(g)=g^2/4\pi$. From Eq.~(\ref{rg2}), one can check that our result
Eq.~(\ref{final}) is independent of the scale $M$.

We have approximated the interaction by a delta-function which has zero range.
In real systems, however,  the interactions have a finite range $a$ and
so $1/a$ provides a natural ultraviolet cutoff.

\section{$T$-matrix Approximation}
In this section we employ the ladder or $T$-matrix approximation, which
takes an infinite number of loops into account. 

\begin{figure}[htb]
\begin{center}
\begin{picture}(0,0)%
\epsffile{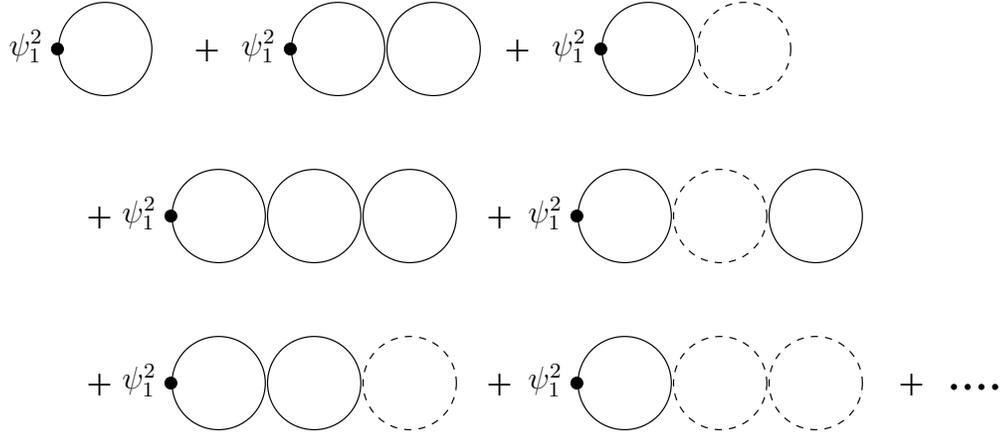}%
\end{picture}%
\setlength{\unitlength}{1973sp}%
\begingroup\makeatletter\ifx\SetFigFont\undefined%
\gdef\SetFigFont#1#2#3#4#5{%
  \reset@font\fontsize{#1}{#2pt}%
  \fontfamily{#3}\fontseries{#4}\fontshape{#5}%
  \selectfont}%
\fi\endgroup%
\begin{picture}(11775,5400)(676,-5761)
\put(676,-1036){\makebox(0,0)[lb]{\smash{\SetFigFont{12}{14.4}{\familydefault}{\mddefault}{\updefault}$\psi_{1}^{2}$}}}
\put(7501,-1036){\makebox(0,0)[lb]{\smash{\SetFigFont{12}{14.4}{\familydefault}{\mddefault}{\updefault}$\psi_{1}^{2}$}}}
\put(7201,-3136){\makebox(0,0)[lb]{\smash{\SetFigFont{12}{14.4}{\familydefault}{\mddefault}{\updefault}$\psi_{1}^{2}$}}}
\put(2101,-3136){\makebox(0,0)[lb]{\smash{\SetFigFont{12}{14.4}{\familydefault}{\mddefault}{\updefault}$\psi_{1}^{2}$}}}
\put(2101,-5236){\makebox(0,0)[lb]{\smash{\SetFigFont{12}{14.4}{\familydefault}{\mddefault}{\updefault}$\psi_{1}^{2}$}}}
\put(7201,-5236){\makebox(0,0)[lb]{\smash{\SetFigFont{12}{14.4}{\familydefault}{\mddefault}{\updefault}$\psi_{1}^{2}$}}}
\put(3601,-1036){\makebox(0,0)[lb]{\smash{\SetFigFont{12}{14.4}{\familydefault}{\mddefault}{\updefault}$\psi_{1}^{2}$}}}
\end{picture}

\vspace{4mm}
\caption[a]{Series of diagrams that are summed to give the leading 
correction to the mean-field term $g\rho\sqrt{\rho}$ in Eq.~(\ref{final}).}
\label{ladder}
\end{center}
\end{figure}

The diagrams we are summing and that give the leading correction to the
mean-field term $g\rho\sqrt{\rho}$ in Eq.~(\ref{final}) are shown in 
Fig.~\ref{ladder}. 
The diagrams are not the conventional ladder diagrams
since we use two real fields instead of a single
complex field. The final result, however, is the same.
The correction to the gradient term 
in Eq.~(\ref{final}) is obtained by 
summing the same diagrams with insertions of sources $X$, $Y$, and their
derivatives in the leftmost loop in every diagram.

The ladder diagrams are summed by solving an integral equation for 
the two-body $T$-matrix~\cite{schiff,popov,henk}.
In two dimensions, the summation
amounts to replacing the bare
interaction $g$ by an effective interaction $8\pi\log(2g\rho a^2)^{-1}$,
which is density dependent~\cite{schiff,popov,henk}. 
In the dilute gas limit, this expansion parameter is small, and 
Eq.~(\ref{final}) is replaced by 
\bqa
\label{identi}
\left[
\nabla^2+\mu-V
\right]\sqrt{\rho}-8\pi|\log{(2g\rho a^2)}|^{-1}\rho\sqrt{\rho}
-{2\over3}
|\log{(2g\rho a^2)}|
^{-1}\nabla^2\sqrt{\rho}
=0\;.
\eqa
This is the main result of the present paper. 
The second term was
obtained in Refs.~\cite{lieb1,lieb2,kolo} and follows from the
corrections to the ground state energy of a homogeneous Bose gas 
calculated by Schick~\cite{schiff}.
The third term which is a correction to the gradient term is a new result.

Up to corrections that are suppressed by powers of the small parameter
$\log(2g\rho a^2)^{-1}$, Eq.~(\ref{identi}) can 
be derived from an energy functional 
$E[\Phi]$, where
$|\Phi|$, is identified with $\sqrt{\rho}$:
\bqa\nonumber
E[\Phi]&=&
{\hbar^2\over2m}\int d^2x\Bigg\{
\left[1
-{2\over3}
|\log{(2g|\Phi|^2a^2)}|^{-1}\right]|\nabla\Phi|^2
+{2m\over\hbar^2}\left[
V-\mu\right]|\Phi|^2\\ 
\label{ef}
&&+{4\pi}|\log{(2g|\Phi|^2 a^2)}|^{-1}|\Phi|^4
\Bigg\}\;.
\eqa
The energy functional Eq.~(\ref{ef}) generalizes the energy
functional in Refs.~\cite{lieb1,kolo}
to include a gradient term.

\section{Summary}
In this paper, we have computed the leading quantum corrections to the
Gross-Pitaevskii equation for a trapped Bose gas in two spatial dimensions.
The method involves the truncation of two systematic expansions. The first
is the quantum loop expansion which is an expansion in powers of 
the coupling constant $g$, and the second is a gradient expansion around the
Thomas-Fermi limit. The result Eq.~(\ref{final}) includes all leading order 
quantum corrections to second order in the gradient expansion.

The summation of the ladder diagrams changes the effective expansion
parameter from $g$ to $8\pi\log(2g\rho a^2)^{-1}$.
The resulting equation~(\ref{identi})
includes a correction term that can be 
derived from the corrections to the ground state energy density
of a homogeneous Bose gas. It also includes a new term which
is a gradient correction to the mean field equation.

\section{Acknowledgments}
This work was supported by the Stichting Fundamenteel Onderzoek der Materie
(FOM), which is supported by the Nederlandse Organisatie voor Wetenschapplijk
Onderzoek (NWO). The authors would like to thank H. T. C. Stoof for 
useful discussions.
\appendix
\renewcommand{\theequation}{\thesection.\arabic{equation}}
\setcounter{equation}{0}
\section{Formulas}
The loop integrals that appear in our calculations involve integrations
over the energy $\omega$ and the spatial momentum ${\bf k}$.
The energy integrals are evaluated using contour integration. 
The specific integrals needed are
\bqa
\label{e1}
\int{d\omega\over2\pi}{1\over(\omega^2-\epsilon^2+i\epsilon)^n}&=&
i(-1)^{n+1}{(-1)\cdot1\cdot3...(2n-3)\over2^n(n-1)!}{1\over\epsilon^{2n-1}}
\;,\\
\int{d\omega\over2\pi}{\omega^2\over(\omega^2-\epsilon^2+i\epsilon)^{n+1}}&=&
i(-1)^{n+1}{(-1)\cdot1\cdot3...(2n-3)\over2^{n+1}n!}{1\over\epsilon^{2n-1}}\;.
\eqa
The momentum integrals are evaluated 
using dimensional regularization in $d=2-2\epsilon$ 
dimensions. 
Some of the integrals are infrared divergent or ultraviolet divergent or
both.
They can be written in terms of integral $I_{m,n}$, which is defined by
\bqa
\label{idef}
I_{m,n}=\left({e^{\gamma}M^2\over4\pi}\right)^{\epsilon}
\int{d^dk\over(2\pi)^d}{k^{2m}\over k^n(k^2+\Lambda^2)^{n/2}}\;.
\eqa
Here, $M$ is a renormalization scale that ensures that $I_{m,n}$ has the
canonical dimension also for $d\neq2$.
$\gamma\approx0.5772$ is the Euler-Mascheroni constant. 
With dimensional regularization, $I_{m,n}$ is given by the formula
\bqa
I_{m,n}={\Omega_d\over(2\pi)^d}
\left({e^{\gamma}M^2\over4\pi}\right)^{\epsilon}
\Lambda^{d+2m-2n}
{\Gamma({{d-n\over2}+m})\Gamma(n-m-{d\over2})\over2\Gamma({n\over2})}\;,
\eqa
where $\Omega_d=2\pi^{d/2}/\Gamma[d/2]$ is the 
area of the $d$-dimensional sphere.

The integrals $I_{m,n}$ satisfy the relations
\bqa
\left(d+2m-n\right)I_{m,n}&=&nI_{m+2,n+2}\;,\\
\Lambda^2I_{m,n}&=&I_{m-1,n-2}-I_{m+1,n}\;.
\eqa
The first relation follows from integration by parts, while the second
is simply an algebraic relation.
\section{Coefficients}
The coefficients needed to calculate 
$\langle\psi_1^2\rangle$ and $\langle\psi_2^2\rangle$
in Eqs.~(\ref{p1}) and~(\ref{p2})
were calculated and listed 
in Ref.~\cite{eric}. For completeness, we list them below
\bqa
a_0&=&{1\over2}I_{1,1}\;,\\
a_1&=&{1\over4}I_{2,3}\;,\\
a_2&=&-{1\over4}I_{0,1}\;,\\
a_3&=&{1\over16d}\left[-10I_{5,7}+13I_{3,5}-I_{1,3}\right]\;,\\
a_4&=&{3\over16}I_{3,5}\;,\\
a_5&=&{1\over64d}\left[35I_{6,9}-10I_{4,7}+3I_{2,5}\right]\;,\\
b_0&=&{1\over2}I_{-1,-1}\;,\\
b_1&=&-{1\over4}I_{0,1}\;,\\
b_2&=&{1\over4}I_{-2,-1}\;,\\
b_3&=&{1\over16d}\left[2I_{3,5}-3I_{1,3}-I_{-1,1}\right]\;,\\
b_4&=&-{1\over16}I_{1,3}\;,\\
b_5&=&-{5\over64d}\left[I_{4,7}+2I_{2,5}+I_{0,3}\right]\;.
\eqa
The factor of $d$ in the denominators in $a_3$, $a_5$, $b_3$, and
$b_5$ arises from averaging over angles in the momentum integrals.

Finally, we outline how to calculate a one-loop diagram
explicitly. For simplicitly, we compute the leading coefficient $a_0$ in the
gradient expansion for the expectation value 
$\langle\psi_1^2\rangle$ in Eq.~(\ref{p1}).
The diagram is shown in Fig.~\ref{ex}, and it reads
\bqa
a_0=\left({e^{\gamma}M^2\over4\pi}\right)^{\epsilon}\int{d\omega\over2\pi}\int{d^dk\over(2\pi)^d}
{ik^2\over\omega^2-\epsilon^2+i\epsilon}\;.
\eqa
After integrating over the energy $\omega$, using Eq.~(\ref{e1}), we obtain
\bqa\nonumber
a_0&=&{1\over2}\left({e^{\gamma}M^2\over4\pi}\right)^{\epsilon}\int{d^dk\over(2\pi)^d}{k^2\over\epsilon}\\ 
&=&{1\over2}I_{1,1}\;.
\eqa


\begin{thebibliography}{99}
\bibitem{bec1} M. H. Anderson, J. R. Ensher, M. R. Matthews, C. E. Wieman
and E. A. Cornell, Science {\bf 269}, 198 (1995).
\bibitem{bec2}
K. B. Davis, M. O. Mewes, M. R. Andrews, N. J. van Druten,
D. S. Durfee, D. M. Kurn and W. Ketterle, Phys. Rev. Lett {\bf 75}, 3969 (1995).
\bibitem{bec3}
C. C. Bradley, C. A. Sackett, J. J. Tollet, and R. G. Hulet, 
Phys. Rev. Lett. {\bf 75}, 1687 (1995).
\bibitem{string} F. Dalfovo, S. Giorgini, L. P. Pitaevskii, and S. Stringari, Rev. Mod. Phys. {\bf 71}, 463 (1999).
\bibitem{corny}
L. Cornish, N. R. Claussen, J. L. Roberts, E. A. Cornell, and C. E. Wieman, cond-mat/0004290.

\bibitem{eric}E. Braaten and A. Nieto, Phys. Rev. {\bf B 56}, 14745 (1998).
\bibitem{jenseric} J. O. Andersen and E. Braaten, Phys. Rev. {\bf A 60}, 2330
(1999).
\bibitem{qc}D. Blume and C. H. Greene, cond-mat/0009220.
\bibitem{hohen}P. C. Hohenberg, Phys. Rev. {\bf 158}, 383 (1967).
\bibitem{schiff}M. Schick, Phys. Rev. {\bf A 3}, 1067 (1971).
\bibitem{fis}D. S. Fisher and P. C. Hohenberg, Phys. Rev. {\bf B 37}, 4936 (1988).
\bibitem{rg2}E. B. Kolomeisky, and J. P. Straley, Phys. Rev. {\bf B 46}, 
11749 (1992). 

\bibitem{lieb1}E. H. Lieb and J. Yngvason, math-ph/0002014.
\bibitem{lieb2}E. H. Lieb, R. Seiringer, and J. Yngvason, cond-mat/0005026.
\bibitem{klep}V. Bagnato and D. Kleppner, Phys. Rev. {\bf A 44}, 7439 (1991).

\bibitem{jason}T-L. Ho and M. Ma, J. Low Temp. Phys. {\bf 115}, 61 (1999).

\bibitem{petrov}D. S. Petrov, M. Holzmann, and G. V. Shlyapnikov,
Phys. Rev. Lett. {\bf 84}, 2551 (2000).
\bibitem{torhaarek}T. Haugset and H. Haugerud, Phys. Rev. {\bf A 57}, 3809 (1998).\bibitem{adhi}S. K. Adhikari, Phys. Lett. {\bf A 265}, 91 (2000).
\bibitem{kolo}
E. B. Kolomeisky, T. J. Newman, J. P. Straley, and X. Qi, cond-mat/0002282.









\bibitem{berg}O. Bergman, Phys. Rev. {\bf D 46}, 5474 (1992). 
\bibitem{loz}G. Lozano, Phys. Lett. {\bf B 283}, 70 (1992).
\bibitem{finn}T. Haugset and F. Ravndal, Phys. Rev. {\bf D 49}, 4299 (1994).
\bibitem{popov}V. N. Popov, {\it Functional Integrals and collective
excitations,} (Cambridge University Press 1987.).

\bibitem{henk}H. T. C. Stoof and M. Bijlsma, Phys. Rev. {\bf E}, 939 (1993).
\end{thebibliography}
\end{document}